\documentclass[conference]{IEEEtran}
\usepackage{graphicx}
\usepackage{amssymb}
\usepackage{amsmath}
\usepackage{flexisym}
\usepackage[noadjust]{cite}
\usepackage{float}
\usepackage{subcaption}
\usepackage[font=small]{caption} 
\begin{document}
%
\title{Optimal Traffic Splitting Policy in LTE-based Heterogeneous Network}
\author{\IEEEauthorblockN{Pradnya Kiri Taksande,
Arghyadip Roy, 
Abhay Karandikar}
\IEEEauthorblockA{Department of Electrical Engineering,
Indian Institute of Technology Bombay, India 400076\\
Email: \{pragnyakiri,arghyadip,karandi\}@ee.iitb.ac.in}
}
\maketitle
%
\begin{abstract}
Dual Connectivity (DC) is a technique proposed to address the problem of increased handovers in heterogeneous networks. In DC, a foreground User Equipment (UE) with multiple transceivers has a possibility to connect to a Macro eNodeB (MeNB) and a Small cell eNodeB (SeNB) simultaneously. In downlink split bearer architecture of DC, a data radio bearer at MeNB gets divided into two; one part is forwarded to the SeNB through a non-ideal backhaul link to the UE, and the other part is forwarded by the MeNB. This may lead to an increase in the total delay at the UE since different packets corresponding to a single transmission may incur varying amounts of delays in the two different paths. Since the resources in the MeNB are shared by background legacy users and foreground users, DC may increase the blocking probability of background users. Moreover, single connectivity to the small cell may increase the blocking probability of foreground users. Therefore, we target to minimize the average delay of the system subject to a constraint on the blocking probability of background and foreground users. The optimal policy is computed and observed to contain a threshold structure. The variation of average system delay is studied for changes in different system parameters.
\end{abstract}
%
%
\IEEEpeerreviewmaketitle
\section{Introduction}
\label{sec1}
With an upsurge in the use of smartphones and tablet devices, the mobile data traffic is proliferating. According to \cite{cisco}, the monthly global mobile data traffic is predicted to reach 30.6 exabytes by 2020. The deployment of Heterogeneous Networks (HetNet) comprising of small cells overlaid with ubiquitous macro cells is one of the significant approaches to meet this ever-increasing demand for mobile data traffic. Although the introduction of HetNets is beneficial in many aspects, it leads to an increase in the number of UE handovers and signaling overhead, due to the difference in the coverage areas of small and macro cells. To combat this, 3rd Generation Partnership Project (3GPP) has proposed Control-plane/User-plane split \cite{splitCU,phantom} and the Dual Connectivity (DC) technique as a part of Long Term Evolution (LTE) Release 12 \cite{36842}. In Control-plane/User-plane split, macro cells manage the Control-plane whereas the small cells handle the User-plane. DC allows a User Equipment (UE) with multiple transceivers to simultaneously receive data from both a Macro eNodeB (known as Master eNodeB) and a Small cell eNodeB (known as Secondary eNodeB). In this paper, we study the optimal splitting policy for DC UEs.\par
\begin{figure}[!htb]
\centering
\includegraphics{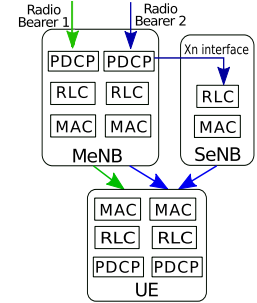}
\caption{Dual Connectivity user plane protocol stack.}
\label{userplanearch}
\end{figure}

We consider the split bearer architecture \cite{36842} of DC, which has a user plane protocol stack as depicted in Figure \ref{userplanearch}.
In this architecture, only the Macro eNodeB (MeNB) has a connection with the Core Network. The MeNB manages the Control-plane of UE, whereas its User-plane can be split between the MeNB and the Small cell eNB (SeNB). The MeNB and SeNB are connected via the Xn interface, which is a non-ideal backhaul link. The data of a radio bearer for a UE arrives from the higher layers at the Packet Data Convergence Protocol (PDCP) layer of MeNB; MeNB then splits it into two parts, as shown by Radio Bearer 2 in the figure. One part is forwarded to the SeNB via the backhaul link, which then transmits to the UE and the other part is transmitted by the MeNB. The aggregation of the split bearer then takes place at the PDCP layer of the UE.\par
A DC-capable UE can use DC to significantly increase its throughput and improve its mobility performance \cite{functionality}. However, there may be considerable delays in the reception of DC traffic at the UE because the first and the last packet corresponding to a single transmission may arrive via two different paths with widely varying delays. The legacy UEs (background UEs) can connect to MeNB only. For the UEs which are capable of DC (foreground UEs), data traffic can be received via MeNB or SeNB or both. Since the resources in the MeNB are shared by background and foreground UEs, DC may increase the blocking probability of background UEs. Single connectivity of foreground UEs with SeNB may bring down the blocking probability of background UEs, by saving the MeNB resources for background UEs. However, it again increases the blocking probability of foreground UEs, since the MeNB resources are not utilized for foreground UEs. Hence, we introduce a constraint on the weighted sum of the blocking probabilities of background and foreground UEs. Our objective is to minimize the average delay of the system subject to a constraint on the blocking probability of
background and foreground UEs.\par
In \cite{wangdual}, the authors propose a flow control algorithm in which SeNB periodically sends data requests to the MeNB, depending on the buffer status at SeNB. In \cite{dts}, the authors propose a downlink traffic scheduling scheme for maximization of the network throughput. \cite{sumrate, propfair} deal with maximizing the data rate of DC users in LTE and multiple-Radio Access Technology (RAT) scenario, respectively. The works \cite{propfair,dts,loaddc} consider throughput as the system metric of interest. However, none of them consider the delay in the system, which requires attention considering the varying network conditions in the two different paths.\par
The authors in \cite{loaddc} propose a split bearer algorithm for video traffic to improve the data rate. In \cite{optimalflowbifur}, the optimal splitting ratio for minimizing the queuing delay in the system is calculated for a single UE. The authors in \cite{delayoptimal} obtain the optimal traffic splitting over multiple Radio Access Technologies (RATs) such that maximum average delay across different RATs is minimized. They, however, do not consider user arrival and departure. Also, in \cite{delayoptimal}, the authors consider the maximization of expected delays in different RATs as the optimization parameter. However, in our work, we deal with expected maximum delay as the system metric which captures the real life scenario better than that by \cite{delayoptimal}. 
\par
Our contribution is twofold. First, we obtain the optimal splitting policy to minimize the average delay in the system subject to a constraint on the blocking probability. Second, we demonstrate the variation of average delay in the system as a function of load in the system and backhaul delay. To the best of our knowledge, this work is the first attempt to present an optimal splitting policy for minimizing the average delay in the system using DC enhancement. \par
The paper has the following organization. In Section \ref{sec2}, we outline the system model. The problem formulation and solution methods are explained in Section \ref{sec3}. The structure of the optimal policy along with some numerical results are described in Section \ref{sec4} followed by conclusion in Section \ref{conc}.
\section{System Model}
\label{sec2}
\begin{figure}[!ht]
	\centering
	\includegraphics[width=8.4cm]{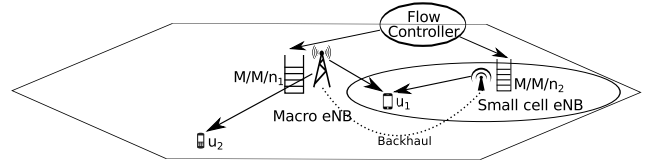}
	\caption{System Model.}
	\label{SysModel}
\end{figure}

We consider a macro cell with a wide coverage area and a small cell situated inside the macro cell as presented in Figure \ref{SysModel}. Let $d$ be the one-way latency of the backhaul link connecting the SeNB with the MeNB. SeNB uses this backhaul link to share its status information with the MeNB, and MeNB uses it to share control/data information with the SeNB. We assume MeNB and SeNB operate at different carrier frequencies. As data traffic over the Internet is bursty in nature, we consider batch arrivals with a random number of packets in a batch. The batch size $G$ follows a discrete probability distribution $ \alpha_i = P(G = i), i = 1,2,\cdots $ with mean batch size $\bar{G}$. The flow controller is situated at the MeNB. It routes the incoming traffic to MeNB or SeNB appropriately. \par
We segregate the UEs into two categories. The legacy UEs which are present in the coverage area of the macro cell and can connect to MeNB only (e.g., $u_2$ in Figure \ref{SysModel}) are categorized as background UEs. The UEs which are present in the coverage area of the small cell and capable of dual connectivity to the MeNB and SeNB (e.g.,$u_1$ in Figure \ref{SysModel}) are categorized as foreground UEs. 
The data traffic streams for these two sets of UEs are each assumed to constitute two Poisson arrival streams with rates $\lambda_1$ and $\lambda_2$, respectively. The service times of a packet in MeNB and SeNB are exponentially distributed with mean $1/\mu_m$ and $1/\mu_s$, respectively. All UEs are assumed to be stationary. \par
In LTE, each eNB is assigned a certain number of resources. We assume each fixed size packet of a batch requires one server, i.e., one resource of an eNB to get served. After all the resources are exhausted, the packets are placed in a queue at the eNB. We assume the queue size is large but finite (say, $N$) for both the systems. After the packets join any of the two systems, the scheduling of packets in both the systems takes place independently of each other. Thus, MeNB and SeNB are modeled as $M/M/n$ queuing systems with First-Come-First-Serve queuing discipline. \par
The background traffic can join the MeNB or get rejected. For the foreground traffic, the flow controller at the PDCP layer of MeNB needs to take an appropriate decision regarding admission and splitting of traffic between the two systems. We assume that both types of UEs are assigned equal priority while allocating resources. Henceforth, we denote the MeNB system as System M and the SeNB system as System S.\par
%
\subsection{States}
We model the system as a continuous time stochastic process $\{X(t)\}_{t\geqslant0}$ defined on state space $\mathcal{S}$. A state $s \in \mathcal{S}$ is represented as a 3-tuple $(s_1,s_2,k)$, where $s_1$, $s_2$  represent the number of packets in the queue plus the number of packets currently in service in the System M and S, respectively. $k$ takes different values based on the arrival of a batch or departure of a packet. In the case of departure of a packet, $k=0$. If there is a foreground batch arrival of size $G=1,2,...,n$ then $k$ takes values $1,2,...n$, respectively. If there is a background batch arrival of size $G=1,2,...,n$ then $k$ takes values $n+1,n+2,...2n$, respectively. Since the state of the system changes only at the arrival or departure instants, there is no need to consider the state of the system at other points in time. We explain the state space with an example. For maximum batch size $n=2$, $k=1,2$ represent a foreground traffic arrival of batch size $1$ and $2$, respectively. $k=3,4$ represent a background traffic arrival of batch size $1$ and $2$, respectively. Let $n_1$ and $n_2$ represent the number of resources at MeNB and SeNB, respectively. For instance, consider $n_1=5, n_2=5$ and queue size $N=10$. Then, $s_1 \leqslant n_1+N, s_2 \leqslant n_2+N$. Thus, state $s = (3,6,2)$ indicates that there are $3$ packets in the MeNB system and $5$ packets in service ($n_2=5$) plus $1$ packet in the queue of the SeNB system, when a foreground traffic with batch size 2 ($k=2$) has arrived. \par
\subsection{Decision epochs and Actions}
\indent The decision epochs are the time instants at which the controller needs to take a decision, based on the current system state. The decision epochs are the arrival and departure instances. We denote the actions as $a \in \mathcal{A}$, where $\mathcal{A}$ is the action space. At arrival epochs of the background traffic, the action is to either reject or accept the traffic in the MeNB system. At arrival epochs of foreground traffic, the controller's job is to either reject or decide the appropriate fraction of traffic to route through System M, based on the current state of the system $s$. The action space grows as the size of the batch increases. For instance, in case of maximum batch size $n=2$, the action space is as follows:
\begin{equation*}
\mathcal{A}=
\begin{cases}
0, & \text{Do nothing / Block}, \\
1, & \text{Accept $G$ packets in S},\\
2, & \text{Accept 1 packet in M and $G-1$ packets in S},\\
3, & \text{Accept 2 packets in M and $G-2$ packets in S},\\
\end{cases}
\end{equation*}
with the restriction that $a=0$ is the only feasible decision when there is no room left for any packet. For instance, in the event of departure ($k=0$), the only valid action is $a=0$. For the event of foreground arrival with batch size $G=1$ ($k=1$), the actions $a \in \{0,1,2\} $ are valid, since 1 packet can join the System M ($a=2$) or System S ($a=1$) or it can be rejected ($a=0$). In the event of a foreground batch arrival of size $G=2$ ($k=2$), the actions $a \in \{0,1,2,3\}$ are valid since the 2 packets have the option of joining the System S ($a=1$) or joining the System M ($a=3$) or splitting with 1 packet each joining System M and S ($a=2$) or getting rejected ($a=0$). In the event of background batch arrival of size $G=1,2$ ($k=3,4$), the actions $a \in \{0,2\}$ and $a \in \{0,3\}$, respectively are the only valid actions since background traffic can either be rejected ($a=0$) or made to join the System M ($a=2,3$).
\subsection{Transition probabilities}
At each decision epoch, the controller takes an action $a \in \mathcal{A}$ depending on the state of the system $s$. Depending on the state and action taken, the system moves to another state with a finite probability. Let $T_{ss\textprime}(a)$ denote the transition probability from state $s$ to state $s\textprime$ under the action $a$. Denote $\nu(s_1,s_2)$ as the sum of arrival and departure rates, when the current state is $s = (s_1,s_2,k)$,
\begin{equation}
\label{nu}
\nu(s_1,s_2) = \lambda_1 + \lambda_2 + \min(s_1,n_1)\mu_m + \min(s_2,n_2)\mu_s.
\end{equation}
Note that $\nu(s_1,s_2)$ is independent of $k$.\par 
The transition probabilities from state $s = (s_1,s_2,k)$ to state $s\textprime$ under action $a$ are given by:
 \begin{equation}
    T_{ss\textprime}(s) = 
    \begin{cases}
      \frac{s'_1\mu_m}{\nu(s'_1,s'_2)}, & \text{$s\textprime = (s'_1-1,s'_2,0)$, } \\
      \frac{s'_2\mu_s}{\nu(s'_1,s'_2)}, & \text{$s\textprime = (s'_1,s'_2-1,0)$, } \\
       \frac{\lambda_1 \alpha_1}{\nu(s'_1,s'_2)}, & \text{$s\textprime = (s'_1,s'_2,1)$, } \\
       \frac{\lambda_1 \alpha_2}{\nu(s'_1,s'_2)}, & \text{$s\textprime = (s'_1,s'_2,2)$, } \\
        \frac{\lambda_2 \alpha_1}{\nu(s'_1,s'_2)}, & \text{$s\textprime = (s'_1,s'_2,3)$, } \\
       \frac{\lambda_2 \alpha_2}{\nu(s'_1,s'_2)}, & \text{$s\textprime = (s'_1,s'_2,4)$, } \\
        \end{cases}
 \end{equation}
where $\nu(s_1,s_2)$ is given by (\ref{nu}).
Given the current state $s=(s_1,s_2,k)$ and action $a$, the next state $s'=(s'_1,s'_2,k)$ takes values as tabulated in Table \ref{TPM}.
\begin{table}[!htb]
\caption{State Transition Table}
\centering
\begin{tabular}{|l|l|}
    \hline
        \boldmath{$(k,a)$} & \boldmath{$s'=(s'_1,s'_2)$}\\
    \hline
    ($0,0$) & ($s_1,s_2$)\\
        \hline
    (\{$1,2$\},$1$) & ($s_1,s_2+G$)\\        
        \hline
    ($\{1,2,3\},2$) & ($s_1+1,s_2+G-1$)\\        
        \hline
    ($\{2,4\},3$) & ($s_1+2,s_2+G-2$)\\ 
    \hline
\end{tabular}
\label{TPM}
\end{table}
\subsection{Cost function}
Let $c(s,a)$ denote the cost incurred when the system is in state $s = (s_1,s_2,k)$ and action $a \in \mathcal{A}$ is taken. We define this cost as the expected delay encountered by the arriving batch of packets. Since a batch consists of many packets, the delay of a batch is the response time of the last packet of the batch. Thus, the cost function $c(s,a)$ is the expected response time of the last packet of the arriving batch and is given by,
\begin{equation}
\label{costfn}
 c(s,a) = \mathbb{E} \{ \max\{R_m(s_1),R_s(s_2)\} \},
\end{equation}
where $R_m(s_1)$ and $R_s(s_2)$ denote the response times of the arriving batch in System M and System S, respectively. The response time of a packet is the summation of queuing delay and service time. For instance, if $s=(2,2,2)$ and action $a=2$ is chosen, then $c(s,2) = \mathbb{E} \{\max\{R_m(2),R_s(2)\} \}$. Suppose number of resources is $n_1=5$, $n_2=2$ and queue size $N=5$. Then $R_m(2) \sim exp(1/\mu_m)$. This is because the action $a=2$ will add 1 packet in System M, which has 2 resources occupied out of 5. So this packet will be served in time which is exponentially distributed with parameter $\mu_m$. Also, $R_s(2) = X_1+X_2+d$, where $X_1 \sim exp(1/2\mu_s)$ and $X_2 \sim exp(1/\mu_s)$ and $d$ is the latency of the backhaul link. The action $a=2$ adds 1 packet in System S with 2 packets already in service in the system. The current packet has to wait for $X_1$ time since all resources are occupied ($n_2=2$ and $s_2=2$); then it gets serviced in $exp(1/\mu_s)$ time.\par
Minimization of the average delay of the system may, however, lead to blocking of both background and foreground traffic. The function $b_{b}(s,a) $ ($b_{f}(s,a)$) is defined as  a binary indicator that is set to 1 if the background (foreground) arrival is blocked and to 0 otherwise. The parameter $\delta$, $0 \leqslant \delta \leqslant 1$, which decides how much weight is to be assigned to the blocking probability of each traffic type, depends on the choice of the service provider. We define the blocking cost as the weighted sum of background and foreground traffic blocking cost,
\begin{equation}
\label{block}
 b(s,a) =  \delta b_{b}(s,a) + (1-\delta) b_{f}(s,a).
\end{equation}
\section{Problem Formulation}
\label{sec3}
We aim to split the foreground traffic optimally among the two available paths to minimize the average delay in the system. However, the foreground dual connectivity traffic may use up resources of both the systems M and S, and sufficient resources may not be available for background traffic, which can connect to the System M only. Hence, a constraint on the blocking probability of background single connectivity traffic may be required. However, due to sharing of resources in the MeNB between the two types of UEs, foreground UEs may be forced to move to System S or even blocked. This again increases the blocking probability of foreground traffic. Therefore, we introduce a constraint on the weighted sum of blocking probabilities of background and foreground traffic. Thus, our objective is to minimize the average delay in the system subject to a constraint on the total blocking probability. Since, the times between the decision epochs are random, this leads to the formulation of a constrained Semi-Markov Decision Problem (SMDP).
\subsection{Formulation as Constrained Markov Decision Process (CMDP)}
The average cost criterion is considered as the performance criteria in this work. Let $\mathrm{\Pi}$ be the set of stationary policies. We assume that the Markov chains associated with these policies have no two disjoint closed sets,i.e., the Markov chains are unichain. Let $C(t)$ and $B(t)$ be the total delay and blocking incurred up to time $t$ ($t \geqslant 0$), respectively. The time-averaged delay and blocking can be expressed as,
\begin{equation}
\label{Cbar}
 \bar{C} = \lim_{t \to \infty} \frac{1}{t} \mathbb{E}_{\pi} [C(t)],
 \end{equation}
  and,
\begin{equation}
\label{Bbar}
 \bar{B} = \lim_{t \to \infty} \frac{1}{t} \mathbb{E}_{\pi} [B(t)],
\end{equation}
respectively, where $\mathbb{E}_{\pi}$ is the expectation operator under policy $\pi \in \Pi$. Note that the limits in (\ref{Cbar}) and (\ref{Bbar}) exist since we are considering stationary policies. Our objective is to obtain a policy that minimizes $\bar{C}$ subject to a constraint (say, $B_{max}$) on $\bar{B}$.
\begin{equation} \label{CMDP}
 \mbox{ Minimize   } \bar{C} \mbox{  Subject to }  \bar{B} \leq B_{max}.
\end{equation}
It is a constrained MDP problem with average cost and finite state and action spaces. It is widely known that a stationary randomized optimal policy \cite{altman} exists.
\subsection{Uniformization}
The SMDP problem is converted into a discrete-time MDP problem, using the uniformization method \cite{tijms}. We denote the expected time until the next decision epoch, if the action $a$ is chosen in the state $s = (s_1,s_2,k)$ as $\tau(s,a)$. First, choose a number $\tau$ such that $ 0 < \tau < \displaystyle \min_{s,a} \tau(s,a) $. Consider the discrete-time Markov decision model, with same state space and action space as the SMDP model in Section \ref{sec2} and delay cost, blocking cost and transition probabilities modified as:
\begin{equation*}
 \hat{c}(s,a) = c(s,a) / \tau(s,a),
\end{equation*}
\begin{equation*}
 \hat{b}(s,a) = b(s,a) / \tau(s,a) \quad \text{and}
\end{equation*}
\begin{equation*}
   \hat{T}_{ss\textprime}(a) = 
    \begin{cases}
     (\tau / \tau(s,a)) T_{ss\textprime}(a), & \text{$s \neq s\textprime$, } \\
        (\tau / \tau(s,a)) T_{ss\textprime}(a) + [1 - (\tau/ \tau(s,a))], & \text{$s = s\textprime$. } \\   
  \end{cases} 
\end{equation*}
This discrete-time Markov decision model has the same form of optimality equation as that of the original SMDP model. 
\subsection{Lagrangian Approach}
The constrained problem (\ref{CMDP}) can be converted into an unconstrained problem by using the Lagrangian approach \cite{altman}. Let us consider Lagrange Multiplier $\beta \geq 0$. Define
\begin{equation*}
 \hat{h}(s,a;\beta) = \hat{c}(s,a) + \beta \hat{b}(s,a).
\end{equation*}
The dynamic programming equation yielding the optimal policy is given by,
\begin{equation*}
 V(s) = \min_{a} {\left\{ \hat{h}(s,a;\beta) + \sum_{s\textprime \in \mathcal{S}} \hat{T}_{ss\textprime}(a) V(s\textprime)  \right\} }.
\end{equation*}
The problem can be solved using the Value Iteration Algorithm (VIA) \cite{puterman} for a fixed value of $\beta$. At a particular value of $\beta = \beta^\ast $, minimum cost is obtained for the constrained problem. This value $\beta^\ast$ can be determined by using the gradient descent algorithm following \cite{roy}. The value of $\beta$ at the $n^{th}$ iteration is given by,
\begin{equation*}
 \beta_{n+1} = \beta_n + \frac{1}{n}(\bar{B}_{n} - B_{max}),
\end{equation*}
where $\bar{B}_{n}$ is the blocking probability obtained using the policy $\pi_{\beta_n}$ at iteration $n$.
For this value of $\beta^\ast$, the optimal policy is a mixture of two stationary policies, which can be determined by small deviation $\epsilon$ of $\beta^\ast$ in both directions. This results in two policies $\pi_{\beta^{^\ast}\!-\epsilon}$ and $\pi_{\beta^{^\ast}\!+\epsilon}$ with associated average blocking probability $\bar{B}_{\beta^{^\ast}\!-\epsilon}$ and $\bar{B}_{\beta^{^\ast}\!+\epsilon}$, respectively. Define a parameter $q$ such that $ q\bar{B}_{\beta^{^\ast}\!-\epsilon} + (1-q)\bar{B}_{\beta^{^\ast}\!+\epsilon} = B_{max}$. The optimal policy $\pi^{\ast}$ of the CMDP is randomized mixture of the two stationary policies ($\pi_{\beta^{^\ast}\!-\epsilon}$ and $\pi_{\beta^{^\ast}\!+\epsilon}$), such that at each decision epoch, the first policy is chosen with probability $q$ and the second policy is chosen with probability ($1-q$). Thus, the optimal policy is given by,
\begin{equation}
\label{pol}
 \pi^{\ast} = q\pi_{\beta^{^\ast}\!-\epsilon} + (1-q)\pi_{\beta^{^\ast}\!+\epsilon}.
\end{equation}
\section{Numerical Results and Analysis}
\label{sec4}
In this section, we analyze the optimal policy obtained by solving (\ref{CMDP}). The parameters used for the computation of the optimal policy are as presented in Table \ref{parameters}. We assume $\mu_s \geqslant \mu_m$ because the achievable rate in the coverage area of a small cell is typically higher than that in a macro cell \cite{loadbal}. Although for the computation purpose, we assume a maximum batch size of two, the analysis presented in Section \ref{sec3} holds for any general batch size. The structure of the policy and the variation of the average delay under different parameters is described in this section. \par
\begin{table}
	\caption{System model Parameters}
	\begin{center}
		\begin{tabular}{|l|l|}
			\hline
			\textbf{Parameter} & \textbf{Value}\\
			\hline
			Batch size Distribution ($\alpha_1, \alpha_2$) & 0.5, 0.5\\
			\hline
			Number of Resources ($n_1$,$n_2$) & 6, 6\\
			\hline
			Batch arrival rates ($\lambda_1$, $\lambda_2$) & 6.67, 1 batches/s\\
			\hline
			Service rates ($\mu_m$, $\mu_s$) & 1, 1.5 $(s)^{-1}$\\
			\hline
			Constraint  ($B_{max}$) & 0.02\\
			\hline
			Blocking probability weight parameter ($\delta$) & 0.5\\
			\hline
		\end{tabular}
	\end{center}
	\label{parameters}
\end{table}
\begin{figure*} 
\centering
\begin{subfigure}[b]{0.23\textwidth}
 \centering
 \includegraphics[width=\textwidth]{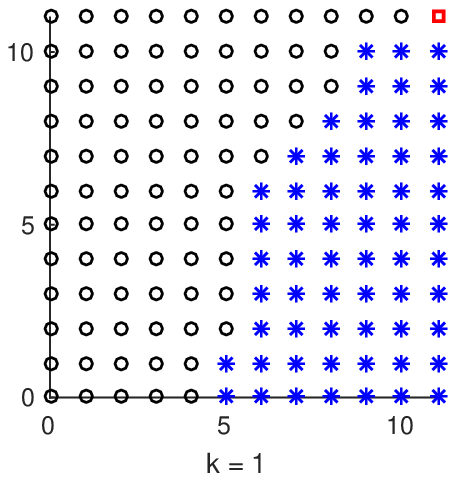}
 \caption{}
 \label{Policy1}
\end{subfigure}%
~ 
\begin{subfigure}[b]{0.23\textwidth}
 \centering
 \includegraphics[width=\textwidth]{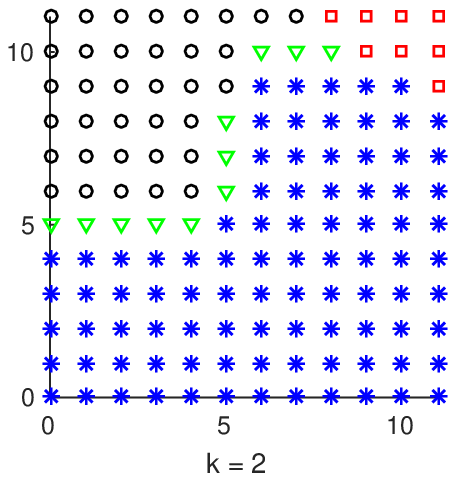}
 \caption{}
 \label{Policy2}
 \end{subfigure}%
~ 
\begin{subfigure}[b]{0.23\textwidth}
\centering
\includegraphics[width=\textwidth]{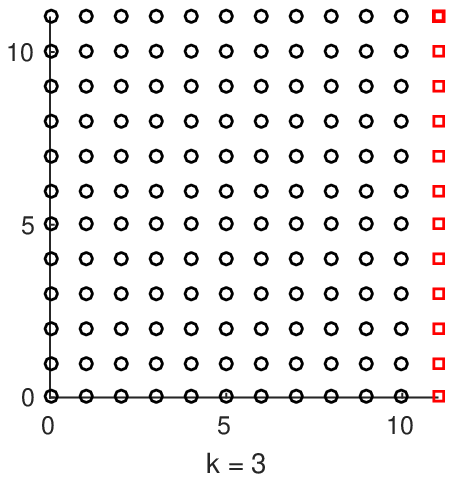}
\caption{}
\label{Policy3}
\end{subfigure}%
~ 
\begin{subfigure}[b]{0.23\textwidth}
 \centering
 \includegraphics[width=\textwidth]{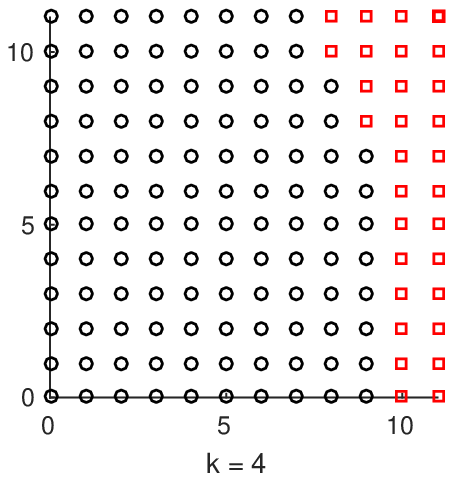}
 \caption{}
 \label{Policy4}  
\end{subfigure}%
\caption{Optimal policy for arrival of foreground traffic with batch of (a) 1 packet ($k=1$), (b) 2 packets ($k=2$), and background traffic with batch of (c) 1 packet ($k=3$) and (d) 2 packets ($k=4$). X and Y axis represent the number of packets (including packets in queue) in System M ($s_1$) and System S ($s_2$), respectively. The squares, asterisks, circles and triangles represent the actions Block ($a=0$), Route to the System S ($a=1$), Route to the System M ($a=3$) and Route to both systems ($a=2$), respectively.}
\end{figure*}
\subsection{Optimal policy structure}
\label{structure}
In this section, the optimal policy obtained by solving the CMDP is outlined. The optimal policy for foreground traffic arrival is illustrated in Figures \ref{Policy1} and \ref{Policy2}. We observe that the optimal policy for foreground traffic arrival with batch size 1 ($k=1$) has a threshold structure. When there are free servers available in System M, the packets are routed to System M. This is because routing to System S incurs an extra backhaul delay, which overrides the benefits achieved from the higher service rate of System S. When the load on System M increases, the queue in System M builds up and the packet at the end of the queue experiences a longer delay in System M. Hence, after a certain point ($s_1=4,s_2=0$), the controller decides to route the arrivals to System S. This is also because the resources of System M need to be reserved for the background traffic, else the blocking probability of the system will increase. If we increase the service rate of System S ($\mu_s$), we observe that the value of this threshold decreases and more traffic is routed through System S. The extra backhaul delay is compensated by the higher service rate of System S. As the number of packets in System S ($s_2$) grows, the choice of system is switched from M to S beyond a threshold. The value of this threshold increases as $s_2$ increases. Again, the optimal action changes to blocking above a certain threshold. Therefore, the policy for foreground traffic arrival with batch size 1 ($k=1$) follows a threshold structure that depends on the number of packets in both the systems and can be expressed as,
\begin{equation}
 \pi^{\ast}(s_1,s_2,1) = 
        \begin{cases}
           1, & \text{$s_1 \le \gamma_1(s_1,s_2)$, } \\
           2, & \text{$\gamma_1(s_1,s_2) < s_1 \le \gamma_2(s_1,s_2)$, } \\
           0, & \text{$s_1 > \gamma_2(s_1,s_2)$, } \\
        \end{cases} 
\end{equation}
where $\gamma_1(s_1,s_2)$ and $\gamma_2(s_1,s_2)$ are thresholds which depend on $s_1$ and $s_2$.\par
For foreground traffic arrival with batch size 2 ($k=2$) (Figure \ref{Policy2}), following a similar argument as in the case of $k=1$, if the System M has servers available ($s_1<5,s_2>5$), then both packets are routed to System M, as shown by circles in the figure. Then as the load on System M increases, the queuing delay increases. Hence, after a threshold on $s_1$, the arrivals are routed to System S to save the resources of System M for the background traffic. When $s_1<5,s_2<5$, all the traffic is routed to System S because the backhaul delay is constant, irrespective of the batch size. Hence, the total delay per packet, which consists of backhaul delay (d) plus response time of the packet, decreases. Therefore, System S is preferred. When there is only one free server in System S, and there are free servers available in System M ($s_2=5, s_1<6$), then one packet is routed to System S, and the other packet is routed to System M, as shown by triangles in Figure \ref{Policy2}. The batch of two packets is split among the two systems to reduce the overall delay in the system; otherwise, the system would suffer an additional queuing delay of 1 packet in System S. Thereafter, the batch is split whenever the delay in System M is nearly the same as that in System S, as shown by the near diagonal structure of triangles in the policy. It is evident from the policy that there exists a threshold, beyond which the batch gets routed to System S. The squares in the figure show that after a threshold on $s_1$, the arrivals are blocked to save resources for background traffic. Thus, the optimal policy for $k=2$ follows a threshold structure and can be expressed as,
\begin{equation}
 \pi^{\ast}(s_1,s_2,2) = 
        \begin{cases}
           1, & \text{$s_1 \le \gamma_3(s_1,s_2)$, } \\
           2, & \text{$\gamma_3(s_1,s_2) < s_1 \le \gamma_4(s_1,s_2)$, } \\
           3, & \text{$\gamma_4(s_1,s_2) < s_1 \le \gamma_5(s_1,s_2)$, } \\
           0, & \text{$s_1 > \gamma_5(s_1,s_2)$, } \\
             \end{cases} 
\end{equation}
where $\gamma_3(s_1,s_2)$, $\gamma_4(s_1,s_2)$ and $\gamma_5(s_1,s_2)$ are thresholds which depend on $s_1$ and $s_2$.\par
The optimal policy for background traffic is illustrated in Figures \ref{Policy3} and \ref{Policy4}. The optimal policy for background traffic arrivals with batch size 1 ($k=3$) is to accept the arrivals in System M and reject beyond a threshold on $s_1$. The optimal policy for $k=3$ can be expressed as,
\begin{equation}
\label{eq1}
 \pi^{\ast}(s_1,s_2,1) = 
        \begin{cases}
           2, & \text{$s_1 \le \gamma_6(s_1,s_2)$, } \\
           0, & \text{$s_1 > \gamma_6(s_1,s_2)$, } \\
        \end{cases} 
\end{equation}
where $\gamma_6(s_1,s_2)$ is a threshold which depends on $s_1$ and $s_2$. Similarly, the optimal policy for background traffic arrival of batch size 2 ($k=4$) is to accept the arrivals in System M and reject them after a threshold on $s_1$. Thus, the optimal policy for $k=4$ follows a threshold structure similar to (\ref{eq1}), where after the threshold the optimum action changes from $a=3$ to $a=0$.
%
%
%
%
\begin{figure*} 
	\centering
	\begin{subfigure}[b]{0.23\textwidth}
		\centering
		\includegraphics[width=\textwidth]{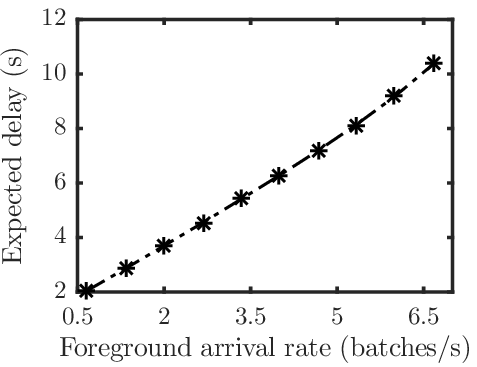}
		\caption{}
		\label{delay_1}
	\end{subfigure}%
	~ 
	\begin{subfigure}[b]{0.23\textwidth}
		\centering
		\includegraphics[width=\textwidth]{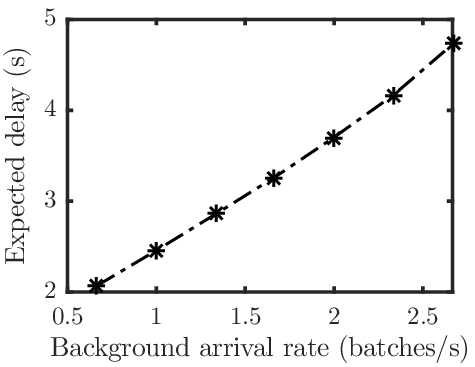}
		\caption{}
		\label{delay_2}
	\end{subfigure}%
	~ 
	\begin{subfigure}[b]{0.23\textwidth}
		\centering
		\includegraphics[width=\textwidth]{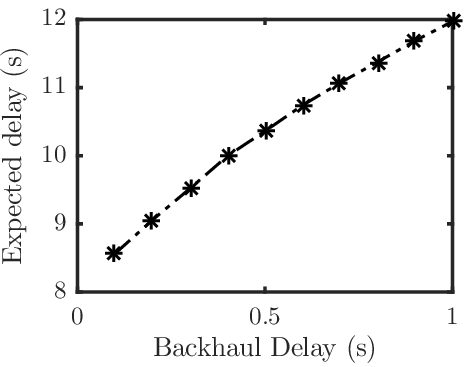}
		\caption{}
		\label{delay_d}
	\end{subfigure}%
	~ 
	\begin{subfigure}[b]{0.23\textwidth}
		\centering
		\includegraphics[width=\textwidth]{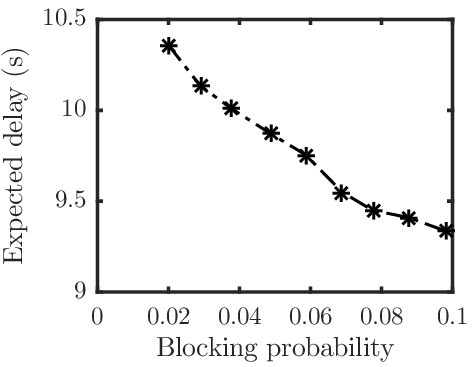}
		\caption{}
		\label{delay_BP}  
	\end{subfigure}%
	\caption{Plots of expected delay vs $\lambda_1,\lambda_2,d$ and blocking probability.}
\end{figure*}
\subsection{Parameter variation}
In this section, we describe the variation of expected delay in the system with the variation of different parameters, following the optimal policy. Figures \ref{delay_1}, \ref{delay_2} and \ref{delay_d} illustrate the expected delay in the system for different values of $\lambda_1, \lambda_2$ and $d$, respectively. In Figure \ref{delay_1}, we vary the foreground arrival rate $\lambda_1$ from 0.67 to 6.67 batches/s with other parameters fixed at $\lambda_2 = 1$ batches/s and backhaul latency $d = 0.5$s. We observe that the expected delay increases steadily with $\lambda_1$. For low values of $\lambda_1$, the Lagrangian Multiplier ($\beta^\ast$) for which the optimal policy is obtained is small. Hence, the difference between the expected delay for CMDP problem and the corresponding unconstrained problem is small. However, as $\lambda_1$ increases, $\beta^\ast$ becomes larger and hence, the rate of increase of expected delay increases. \par
%
In Figure \ref{delay_2}, we keep $\lambda_1=1$ batches/s, $d = 0.5$s and vary background arrival rate $\lambda_2$. As $\lambda_2$ increases, the expected delay in the system rises steadily. For low values of $\lambda_2$, the optimal policy is to route to System M initially and then to System S as explained in Section \ref{structure}. For a higher value of $\lambda_2$, the optimal policy structure remains the same, however, the threshold on $s_1$ changes to a lower value. As background traffic increases, the resources of System M are saved for background traffic and more foreground traffic is routed through System S. \par
In Figure \ref{delay_d}, we keep $\lambda_1,\lambda_2=6.67,1$ batches/s and vary backhaul delay $d$. As $d$ increases, the expected delay in the system rises. For low values of $d$, more foreground traffic is routed to System S reserving the System M for the background traffic. The higher service rate of System S subdues the effect of backhaul delay, and overall delay of the system is low. For high values of $d$, the optimal policy is similar to the policy explained in Section \ref{structure} except for the case $k=2$. For foreground traffic arrival with batch size 2 ($k=2$), the region where the arriving batch is routed to both the systems is increased due to comparable delays in the two systems. The higher value of $d$ is compensated by the higher service rate of System S. The blocking probability is constant at $B_{max}=0.02$ with variation in the parameters $\lambda_1,\lambda_2$ and $d$. \par
In Figure \ref{delay_BP}, we keep $\lambda_1,\lambda_2=6.67,1$ batches/s, $d=0.5$s and vary the blocking probability constraint $B_{max}$. As $B_{max}$ increases, blocking of incoming traffic is allowed more and more which leads to a drop in the delay of the system. We are unable to report all the results due to space constraints.\par
%
%
%
\section{Conclusion}
\label{conc}
In this work, we focus on the problem of varying delays in a split bearer dual connectivity scenario. This is the first work to present an optimal splitting policy using DC enhancement for minimizing the average delay in an LTE-based HetNet subject to a constraint on the blocking probability. The problem is formulated as a constrained SMDP problem, and the optimal policy is observed to contain a threshold structure. We present numerical results which depict the variation of the system delay under different parameter variations. 
%
%
%
\bibliographystyle{ieeetr}
\bibliography{IEEEabrv,MDP_Paper}
\end{document}